\documentclass[twocolumn,numberedappendix]{emulateapj}
\bibpunct[; ]{(}{)}{;}{a}{}{;}
\usepackage{apjfonts}

\usepackage{graphicx}
\usepackage{natbib}

\newcommand{\Msolar}{M${_\odot}$\,}

\shorttitle{Ultraviolet extinction law of 30 Dor}
\shortauthors{De Marchi \& Panagia}

\received{7 Feb 2019}
\accepted{27 Apr 2019}

\begin{document}

\title{Ultraviolet extinction properties of the 30 Dor Nebula \\
 and interpreting observations of starburst clusters\,\altaffilmark{*}}

\author{
Guido De Marchi,\altaffilmark{1}
Nino Panagia,\altaffilmark{2,3}
}

\altaffiltext{1}{European Space Research and Technology Centre,
Keplerlaan 1, 2200 AG Noordwijk, Netherlands; gdemarchi@esa.int}
\altaffiltext{2}{Space Telescope Science Institute, 3700 San Martin
Drive, Baltimore MD 21218, USA; panagia@stsci.edu}
\altaffiltext{3}{Supernova Limited, OYV \#131, Northsound Rd., Virgin Gorda
VG1150, Virgin Islands, UK}

\altaffiltext{{$\star$}}{Based on observations by the
International Ultraviolet Explorer and on observations with the NASA/ESA
{\it Hubble Space Telescope}, obtained at the Space Telescope Science
Institute, which is operated by AURA, Inc., under NASA contract NAS5-26555}

\begin{abstract}

Recent investigation of the extinction law in 30 Dor and the Tarantula
Nebula, at optical and near infrared (NIR) wavelengths, has revealed a
ratio of total to selective extinction $R_V=A_V/E(B-V)$ of about $4.5$.
This indicates a larger fraction of big grains than in the Galactic
diffuse interstellar medium (ISM). Possible origins include coalescence
of small grains, small grain growth, selective destruction of small
grains, and fresh injection of big grains. From a study of the
ultraviolet extinction properties of three massive stars in the 30\,Dor
Nebula (R\,139, R\,140, R\,145), observed with the International
Ultraviolet Explorer (IUE), we show that the excess of big grains does
not come at the expense of small grains, which are still present and
possibly even more abundant. Fresh injection of large grains appears the
dominant mechanism. A process able to naturally account for this in
environments such as the Tarantula Nebula, where formation of massive
stars has been ongoing for over $\sim20$\,Myr, is the  explosion of
massive stars as type-II supernovae (SN). The ensuing change in the
conditions of the ISM is only  temporary, lasting less than $\sim
100$\,Myr, because  shattering and shocks will eventually break and
destroy the bigger grains. However, this is the only time when
star-forming regions are detectable as such in starburst and
high-redshift galaxies and we highlight the complexity inherent in
interpreting observations of star-forming regions in these environments.
If the extinction characteristics are not known properly, any
attempts to derive quantitative physical parameters are bound to fail.

\end{abstract}


\section{Introduction}

The Tarantula Nebula (30\,Dor) in the Large Magellanic Cloud (LMC) is
the nearest extragalactic massive star-forming region. Most of the
energy in this region is produced by NGC\,2070, an OB association
containing at its center the Radcliffe 136 (R\,136) cluster, the closest
example of a massive extragalactic starburst region (Walborn 1991).
R\,136 is  more massive than the supergiant H\,II region NGC\,604 in
M\,33 (e.g. Drissen et al. 1993; Maiz Apellaniz et al. 2004;
Martinez--Galarza et al. 2012) and similar in total mass and energetics
to the super star clusters found in more distant irregular galaxies such
as NGC\,1705 (e.g. Ho \& Filippenko 1996; Tosi et al. 2001) and
NGC\,1569 (e.g. De Marchi et al. 1997; Hunter et al. 2000), or in
starburst galaxies such as M\,82 (e.g. Gallagher \& Smith 1999;
F\"orster Schreiber et al. 2003; Smith et al. 2007). Understanding the
physical properties of these objects and of the environments in which
they are located is the key to properly interpreting also observations
of star formation in high-redshift galaxies (e.g. Vanzella et al. 2017).

A crucial piece of information is a solid characterisation of the
extinction properties in these low-metallicity environments. The
metallicity of the Tarantula Nebula is $Z = 0.007$, a typical value for
the LMC (e.g. Hill et al. 1995; Geha et al. 1998), and corresponds to
$0.5\,Z_\odot$ since $Z_\odot=0.0134$ (Asplund et al. 2009). Thus,
30\,Dor provides an  environment with conditions similar to those
prevailing at redshift  $z \simeq 2$, when star formation in the
universe was at its peak (e.g. Lilly et al. 1996; Madau et al. 1996),
allowing us at the same time to observe and study individual objects
rather than having to resort to their integrated properties. Therefore,
a secure determination of the extinction in the Tarantula Nebula is very
much needed, both in a local and cosmological context.

Recent independent determinations of the extinction properties in and
around 30\,Dor (Maiz Apellaniz et al. 2014; De Marchi \& Panagia 2014; De
Marchi et al. 2016) concur to indicate a value of $R_V=A_V/E(B-V)$
around $4.5$. This implies a larger fraction of big grains than in
the diffuse Galactic interstellar medium (ISM; e.g. Cardelli et al.
1989), similar to what is seen in star-forming  regions in the Milky Way
(MW; e.g. Baade \& Minkowski 1937; Watson \& Costero 2011, and
references therein).

A larger fraction of big grains requires a mechanism that either
selectively adds big grains, or selectively removes small grains by
destroying, growing or co-adding them, or possibly a combination of both.
Regardless of the exact mechanism, the selective removal of small grains
from the mix will have a direct and measurable effect on the shape of
the extinction law at UV wavelengths, resulting in a flatter curve (e.g.
Greenberg 1968)

To probe the UV extinction properties inside the 30 Dor Nebula,
Fitzpatrick \& Savage (1984) studied the spectra of two moderately
reddened Wolf--Rayet (WR) stars close to its centre, namely R\,145 (of
spectral type WN6h+O; Schnurr et al. 2009) and R\,147 (of spectral type
WN5h; Evans et al. 2011; for details on the spectral type
classification of WR stars see, e.g., Crowther 2007). They concluded
that the dense nebular environment permeating 30\,Dor has a reduced
fraction of small grains, compared to the Galactic diffuse ISM, because
the extinction law that they derived does not appear to rise as steeply
as the Galactic law in the far UV.

Their result is based on the comparison of the spectra of R\,145 and
R\,147 with that of R\,144, also a WR object, projected about $4^\prime$
N of the 30\,Dor core and known to be a double-lined spectroscopic
binary of spectral type WN5/6h+WN6 (Sana et al. 2013). In the UV, the
spectral features of R\,144 appear to be compatible with those of R\,145
and R\,147, leading Fitzpatrick \& Savage (1984) to conclude that these
objects might have similar spectral types. However, it is known that the
spectral energy distribution of WR stars at optical and NIR wavelength
is severely affected (flattened) by stellar winds, and current models
are not able to account for the observed differences (e.g. Bonanos et
al. 2009). Therefore, spectral similarities in the UV do not necessarily
extend to longer wavelengths, and in particular they do not guarantee
that the continuum is similar, thus making it difficult to anchor the UV
extinction properties to those in the optical domain and to derive an
internally consistent extinction law.

Instead, armed with the knowledge of a robust optical extinction law now
available for the 30 Dor region from hundreds of red giant stars (De
Marchi \& Panagia 2014; De Marchi et al. 2016), we have re-analised the
UV spectra of both lightly and moderately reddened stars of type O and
WR in 30\,Dor and compared them with atmosphere models following the
``extinction without standards'' technique pioneered by Fitzpatrick \&
Massa (2005; see also Whiteoak 1966). We show that to explain the
observed UV extinction properties inside 30\,Dor there is no need to
invoke a depletion of small grains. Rather, the main difference with
respect to the Galactic diffuse ISM is the presence of a roughly twice
as large relative fraction of big grains and a $\sim 50\,\%$ larger
fraction of small grains too.

The structure of the paper is as follows. In Section\,2 we briefly
discuss the spectroscopic data and their analysis. Section\,3 is devoted
to the comparison between observations and atmosphere models, while in
Section\,4 we address the properties of the derived extinction curve.
The discussion and conclusions follow in Sections\,5 and 6,
respectively.

\begin{figure}
\centering
\resizebox{\hsize}{!}{\includegraphics[trim={3.3cm 0 3.3cm 0},clip]{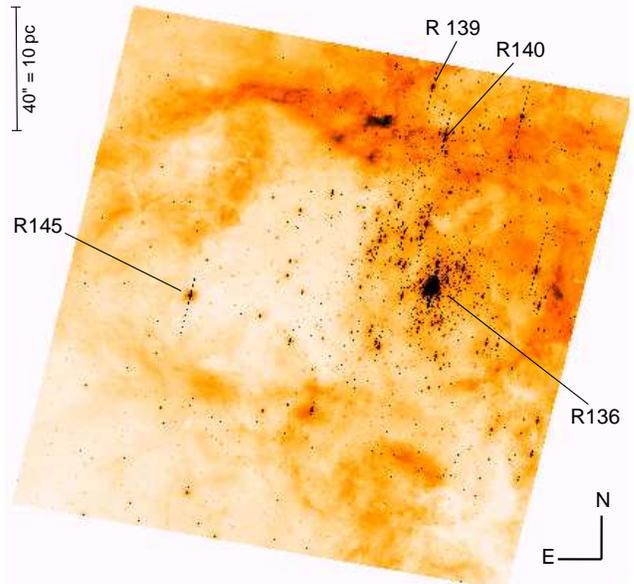}}
\caption{Image of 30\,Dor, covering an area of $\sim 2\farcm7 \times
2\farcm7$ around the R\,136 cluster, with North up and East to the left.
The image, in the F555W band, was obtained with  the WFC\,3 camera on
board the HST (De Marchi et al. 2011). The positions of R\,136, R\,139,
R\,140, and R\,145 are indicated. R\,129 is outside this field, located
$\sim 180$\,pc NNW of R\,136.} 
\label{fig0}
\end{figure}
  
\section{Spectroscopic data}

This paper concentrates on the central $\sim 2\farcm7 \times 2\farcm7$
around the R\,136 cluster (see Figure\,\ref{fig0}). The extinction
of this region has been studied in detail with recent high spatial
resolution {\em Hubble Space Telescope} (HST) multiband photometry in
the $\sim 0.3 - 1.6\,\mu$m wavelength range (De Marchi et al. 2011; De
Marchi, Panagia \& Beccari 2017). Within this region there are three
massive stars whose UV spectra have been obtained with the {\em
International Ultraviolet Explorer} (IUE), namely R\,139, R\,140, and
R\,145. Also the central R\,136 object itself has been observed with the
IUE, but we do not consider it here because multiple massive stars are
known to contribute to the light captured by the IUE $\sim 20\arcsec
\times 10\arcsec$ spectroscopic aperture used in the observations (e.g.
Weigelt \& Baier 1985; De Marchi et al. 1993). In a future work, we will
extend this investigation to other UV spectroscopic observations in the 
wider $\sim 14^\prime \times 12^\prime$ region covered by the {\em
Hubble Tarantula Treasury Project} (Sabbi et al. 2013, 2016), within
which we have already measured the extinction properties at optical and
NIR wavelengths (De Marchi et al. 2016).   

\begin{deluxetable}{llllc} \tablecolumns{4}
\tabletypesize{\footnotesize} 
\tablecaption{IUE observations.
\label{tab1}} 
\tablehead{\colhead{Star} & \colhead{Spectral Type} &
\multicolumn{2}{c}{Observation ID} & 
\colhead{$E(B-V)$} \\[0.05cm] 
\multicolumn{1}{c}{(1)} & \multicolumn{1}{c}{(2)} &
\multicolumn{1}{c}{(3)} & (4) & (5) }
\startdata
R\,129 & ON$9.7$Ia+ & SWP\,5064  & LWR\,4397  &$0.20$ \\
R\,139 & O$6.5$\,Iafc+O6\,Iaf & SWP\,10752 & LWR\,9433  &$0.32$ \\
R\,140 & WN7+WC4/5pec & SWP\,10700 & LWR\,13389 &$0.13$ \\
R\,145 & WN7 & SWP\,14005 & LWR\,10646 &$0.20$
% note line above MUST NOT terminate with \\ for preprint version!
\enddata
\tablecomments{Table columns are as follows: (1) star name;
(2) spectral type; (3) short-wavelength channel spectrum; (4) 
long-wavelength channel spectrum; (5) derived value of $E(B-V)$.}
\end{deluxetable}

R\,139 is a binary system, in which the primary is classified as an
O$6.5$\,Iafc supergiant, while the secondary is an O6\,Iaf supergiant
(Taylor et al. 2011). It is the primary source in this investigation.

R\,140 is a WR system with multiple components, classified as WN6+WC by
Conti (1982) and as WN7+WC4/5pec by Phillips (1982). 

R\,145 was classified as WN6 by Phillips (1982) and by Conti, Leep \&
Perry (1983) and later as WN7 by Moffat (1989), who also identified it
as a binary system. More recently Schnurr et al. (2009) derived the
orbital solution and classified it as WN6h+O. 

A fourth massive star included in our analysis, for comparison, is
R\,129. This blue supergiant was classified as ON$9.7$Ia+ by Walborn
(1977) and it is located just outside of the Tarantula Nebula, to the
NNW of R\,136 at a projected distance of $\sim 180$\,pc (or $\sim
12^\prime$). 

Table\,\ref{tab1} lists the IUE observations that we analyse in this
work. We extracted the calibrated low resolution version of the spectra
from the IUE archive at the Mikulski Archive for Space Telescopes. To
avoid the large photometric uncertainties associated with data collected
through the IUE ``small'' aperture, we restricted our analysis to
spectra secured through the ``large'' $10\arcsec \times 20\arcsec$
aperture. The spectral resolution of these low-dispersion spectra is
about $6 - 7$\,\AA. 

\begin{figure*}
\centering
\resizebox{9.6cm}{6.2cm}{\includegraphics[trim={0 2cm 0.95cm 0},clip]{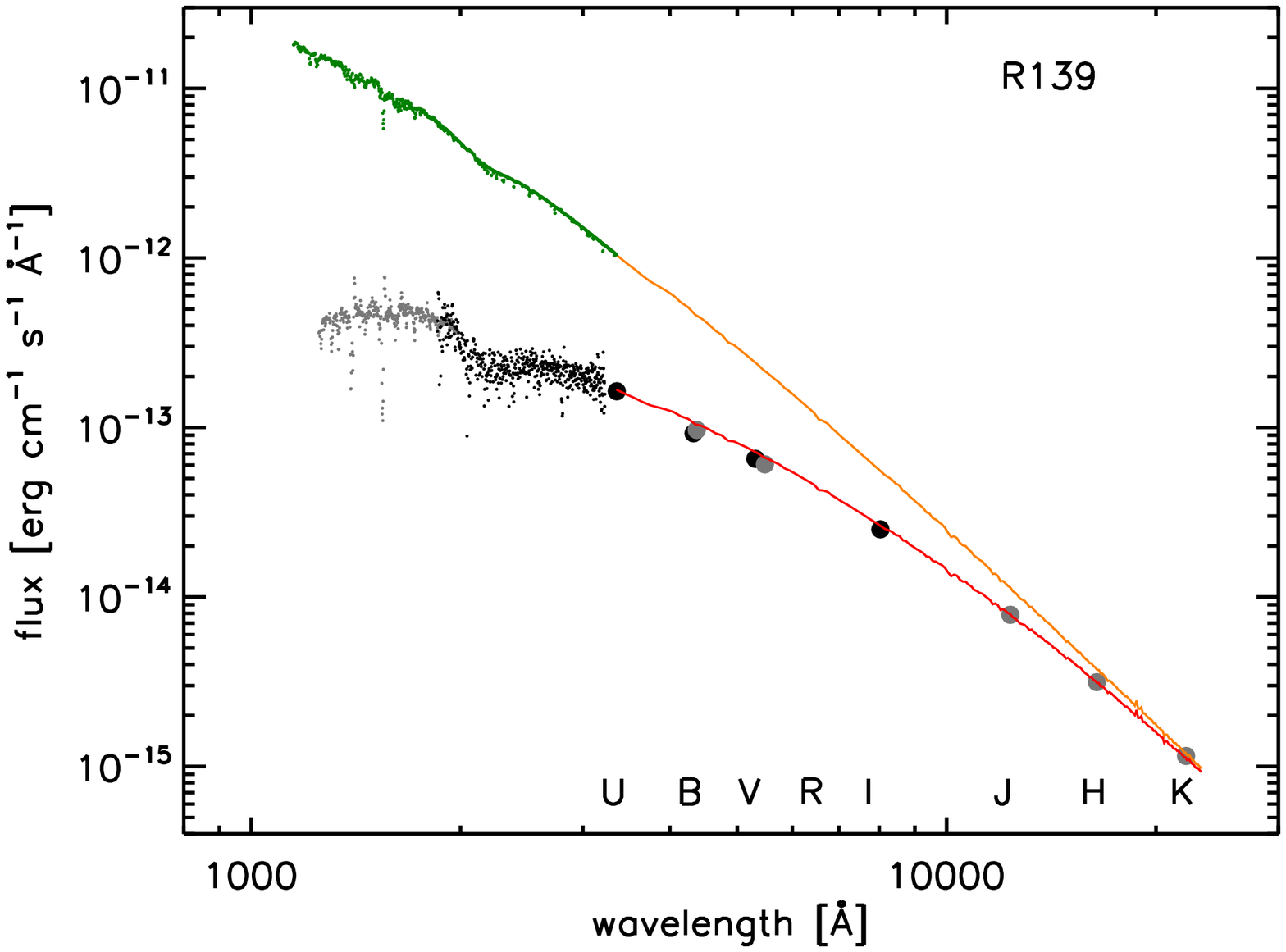}}
\resizebox{0.45\hsize}{6.2cm}{\includegraphics[trim={3.3cm 2cm 0 0},clip]{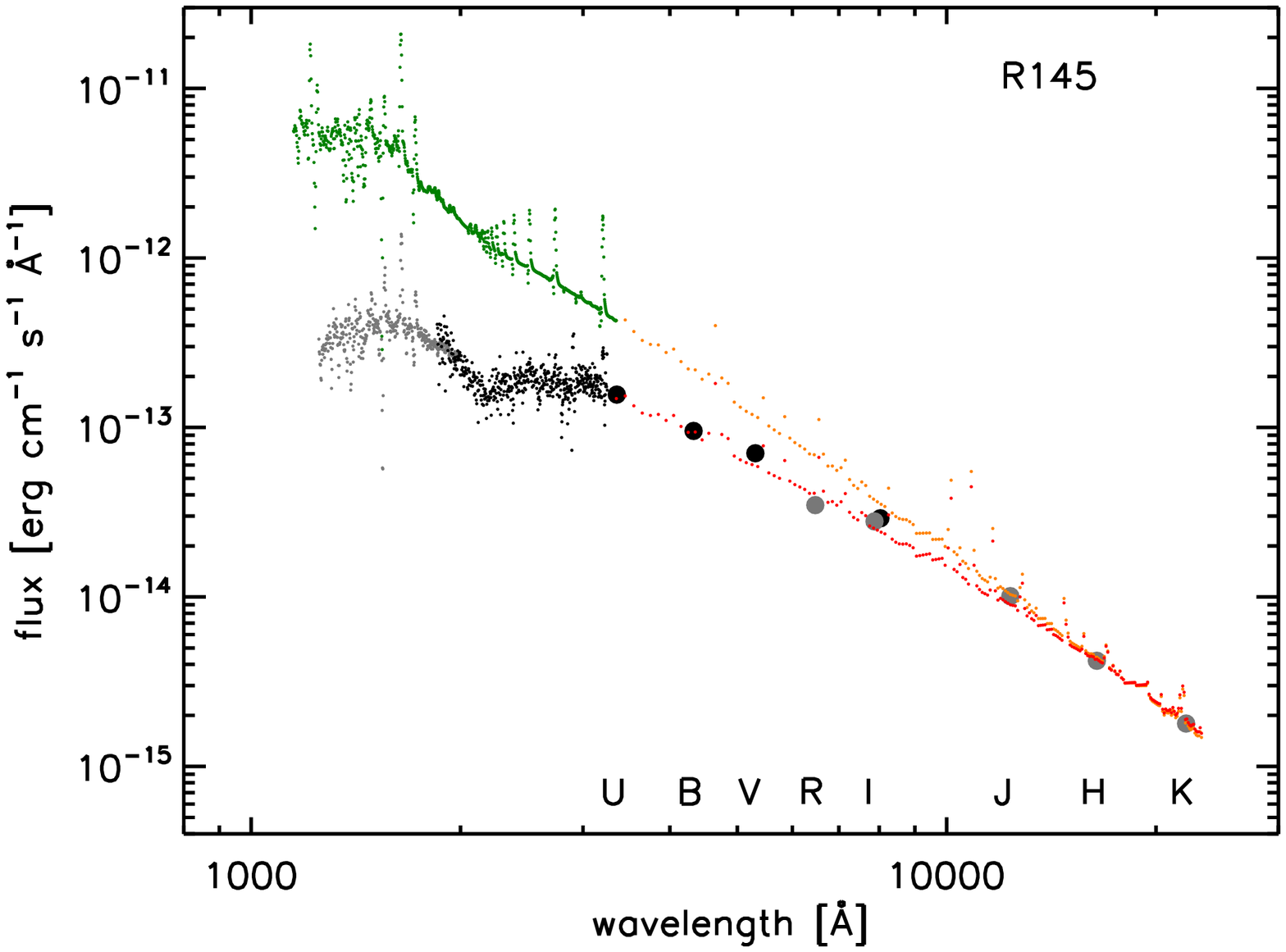}}
\resizebox{9.6cm}{6.7cm}{\includegraphics[trim={0 0 0.95cm 1cm},clip]{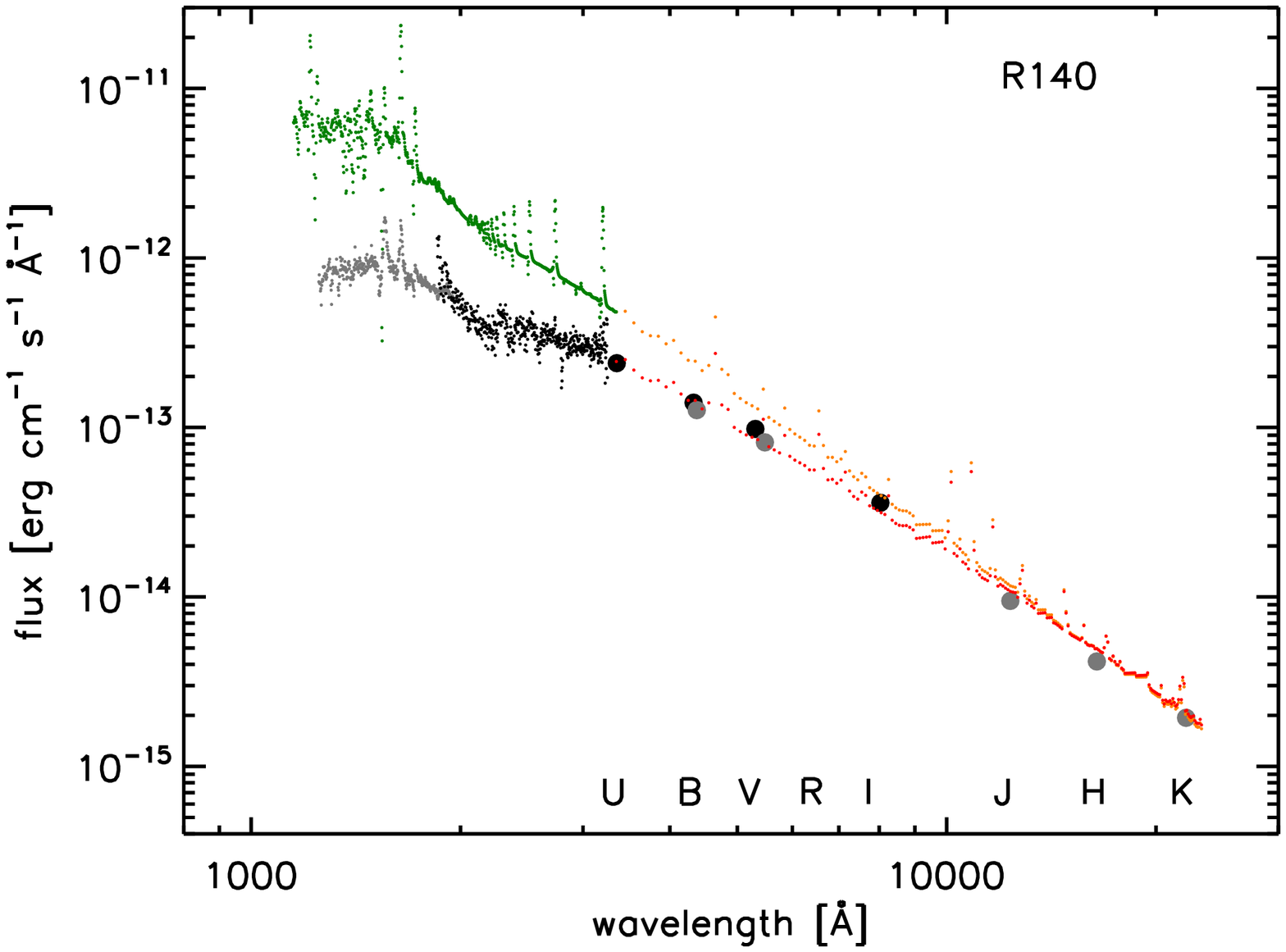}}
\resizebox{0.45\hsize}{6.7cm}{\includegraphics[trim={3.3cm 0 0 1cm},clip]{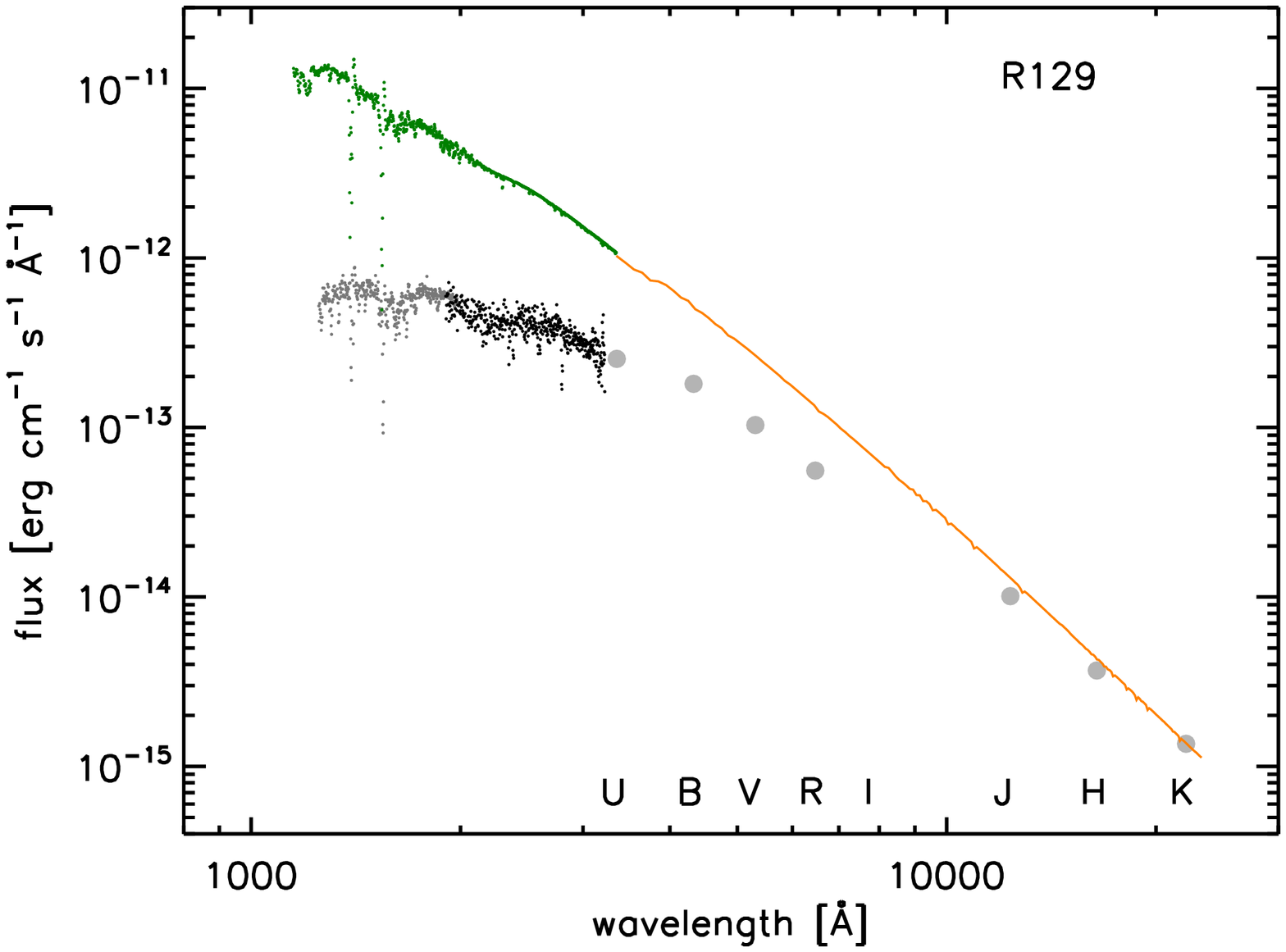}}
\caption{Comparison between observed and theoretical spectra. The IUE
observations, shown by the small grey and black dots, are extended to
visible and NIR wavelengths using broad-band HST photometric
observations from De Marchi et al. (2011; large black dots), as well as
literature broad-band values from SIMBAD (large grey dots; see below for
details). The IUE spectra were normalised to match the photometry in the
$U$ band. The model spectra, shown above each observation, were
convolved with a Gaussian kernel with a full width at half maximum of
6\,\AA, and re-sampled at the same wavelengths of the IUE observations
(small green dots) and every 100\,\AA\ at longer wavelengths (small
yellow dots).  The model spectra already include the effects of the
intervening MW absorption and, for display purposes, are registered with
the photometric observations in the K band. The small red dots show the
best fit to the photometric data obtained by further attenuating the
model spectra according to the 30\,Dor extinction law of De Marchi \&
Panagia (2014), providing in this way the value of $E(B-V)$ for the
stars. Sources of the literature broad-band photometry values are as
follows: for R\,129, Bonanos et al. (2009) for $U$, $V$, and $R$,
Zaritsky et al. (2004) for $B$; for R\,139, Evans et al. (2011) for $B$
and $V$; for R\,140, Howarth (2013) for $B$ and $V$; for R\,145
Zacharias et al. (2013) for $R$ and $I$; for all stars, Cutri et al.
(2003) for $J$, $H$, and $K$. The size of the symbols indicating 
broad-band photometry values exceeds that of the photometric
uncertainties.} 
\label{fig1}
\end{figure*}

The observed IUE spectra are displayed in Figure\,\ref{fig1} as small
grey dots. The large dark dots correspond to the broad-band magnitudes
measured for the same objects by De Marchi et al. (2011) using the  Wide
Field Camera 3 (WFC3) on board the HST, while the large grey dots are
other recent broad-band measurements (see legend) available in the
literature (from SIMBAD)\footnote{The $B$ band photometry given for
R\,129 in SIMBAD, referring to Bonanos et al. (2009) and in turn based
on that originally given by Zaritsky et al. (2004), is incorrect, as
already pointed out by Walborn et al. (2016).}. The IUE spectra were
normalised to match the photometry in the $U$ band obtained by De Marchi
et al. (2011). The scale factors applied to the flux are $0.85$ for
R\,139, $0.80$ for R\,140, $0.93$ for R\,145, and $0.80$ for R\,129.
These values are consistent with the large $10\arcsec \times 20\arcsec$
aperture of IUE collecting also the light of some neighbouring stars,
and with the $\pm 10$\,\% uncertainty on the large aperture's
photometric repeatability (Bohlin et al. 1980; see Bohlin \& Bianchi
2018 for a more recent study). Stellar variability and nebular emission
can also contribute to the apparent mismatch. However, as we will
discuss later, regardless of its specific cause this discrepancy in the
absolute photometric calibration of the IUE data does not affect the
results of this work since we are interested in the overall shape of the
extinction curve in the far UV (FUV) rather than in its detailed
features.

\section{Comparison with spectral models}

To obtain information on the extinction properties of these stars, we
compared the continuum of the observed IUE spectra with that of
theoretical models for similar spectral types. In the case of R\,139, we
selected a model atmosphere from the TLUSTY grid (Lanz \& Hubeny 2003)
{\rm specific for the LMC with effective temperature $T_{\rm
eff}=40\,000$\,K, surface gravity  $\log g=4.0$ and metallicity
$0.5$\,Z$_\odot$ (see Introduction)}. The adopted $T_{\rm eff}$ value is
in line with the O6.5+O6 spectral type of this binary supergiant (e.g.
Panagia 1973). We note that even adopting $T_{\rm eff}=35\,000$\,K or
$45\,000$\,K would have resulted in marginal differences in our
conclusions, less than 1\,\%. For R\,129, given its somewhat later
spectral type (O$9.7$Ia+\,C), we selected from the TLUSTY grid a model
with $T_{\rm eff}=30\,000$\,K, $\log g=4.0$ and metallicity
$0.5$\,Z$_\odot$. Again, using $T_{\rm eff}=27\,000$\,K or $32\,500$\,K
would have resulted in the same conclusions, to within $\sim 6\,\%$
(note that this slightly larger, yet still small, uncertainty compared
to that for the hotter stars is to be expected due to the lower
effective temperature of R\,129). The adopted effective temperature is
consistent with the results of the study of Evans et al. (2004), who
employed the model atmosphere code CMFGEN (Hillier \& Miller 1998) to
investigate the spectral properties of this star and derived $T_{\rm
eff}=27\,500$\,K. 

Since the TLUSTY grid does not include WR model spectra, for the
comparison with R\,140 and  R\,145 we selected a theoretical model from
the Potsdam library (Todt et al.  2015) specific for stars of the WNE
type  and LMC metallicity. The model that best matches the emission
features observed in the IUE spectra is characterised by $T_{\rm
eff}=56\,200$\,K and $\log g=4.2$. 

Although the extinction properties will be derived from the spectral
continuum, a good match also between the observed spectral features and
those predicted by the models is important to ensure a meaningful
comparison. For this reason, the high-resolution model spectra were
first convolved with a Gaussian kernel with a full width at half maximum
of 6\,\AA, comparable with that of the low-resolution IUE observations.
We then re-sampled the resulting spectra at the same wavelengths of the
IUE observations. 

The model spectra are shown by the top curves in each panel in
Figure\,\ref{fig1} and they already include the effects of the
intervening MW absorption along the line of sight, adopting $A_V=0.22$
(Brunet 1975; Fitzpatrick \& Savage 1984) and the standard Galactic
extinction law (Fitzpatrick \& Massa 1990), hence $E(B-V)=0.07$. For
display purposes, the model spectra are registered with the photometric
observations at the red end, in the K band.

Thanks to the robust extinction law recently measured inside 30\,Dor at
optical and NIR wavelengths (Maiz Apellaniz et al. 2014; De Marchi \&
Panagia 2014; De Marchi et al. 2016), from the comparison between the
model spectra and photometry in Figure\,\ref{fig1} we can derive not
only the selective extinction $E(B-V)$, but also the absolute value of
the total extinction $A_V$ towards R\,139, R\,140, and R\,145. We recall
that the 30\,Dor extinction curve corresponds to a value of $R_V=4.5 \pm
0.2$ (De Marchi \& Panagia 2014; De Marchi et al. 2016). The best fits
to the photometric data, indicated by the red curves in
Figure\,\ref{fig1}, correspond respectively to selective extinction
values in the LMC of $E(B-V)=0.32$, $0.15$, and $0.20$, respectively for
R\,139, R\,140, and R\,145. When also the $E(B-V)=0.07$ contribution of
the intervening MW extinction is included, the $E(B-V)$ values become 
$0.39$, $0.22$, and $0.27$ for the three stars,  respectively. The total
combined values of $A_V$ (including the contribution of the MW)
correspond in turn to $A_V=1.66$, $0.90$, and $1.12$ respectively for
R\,139, R\,140, and R\,145. We note in passing that the value of
$E(B-V)=0.16$ quoted by Fitzpatrick \& Savage (1984) for R\,139 appears
incompatible with the recently determined spectral type
(O$6.5$\,Iafc+O6\,Iaf; Taylor et al. 2011) and observed $B-V=0.1$ colour
(Evans et al. 2011) for this object.   

For all stars in our sample we will extend the measurement of the
absolute extinction to near UV and FUV wavelengths simply
by comparing the observed IUE spectra with the model spectra. Indicating
with $F_\lambda^{\rm obs}$ and $F_\lambda^{\rm mod}$ respectively the
observed and model spectrum, the attenuation $A_\lambda$ can be written
as

\begin{equation}
A_\lambda = 2.5 \, \log \frac{F_\lambda^{\rm mod}}{F_\lambda^{\rm
obs}}.
\label{eq1}
\end{equation} 

This operation is particularly straightforward in Figure\,\ref{fig1}. At
UV wavelengths $F_\lambda^{\rm obs}$ and $F_\lambda^{\rm mod}$ have the
same resolution and sampling, so they can be simply divided point by
point. Absolute normalisation is obtained by registering the resulting
curve at the $U$ band.  

R\,129 is located just outside of the 30\,Dor region, in an area of sky
where no direct measurements exist concerning the extinction. Instead of
making assumptions on the possible shape of the extinction curve in this
area, we will use Equation\,\ref{eq1} also to derive the extinction
properties at optical and NIR wavelengths. Indeed, the nearest line of
sight for which extinction has been measured is included in the study
of  Gordon et al. (2003), who measured the properties of the LMC\,2 
supershell surrounding 30\,Dor. However, those lines of sight are
typically  about $0\fdg5$ away from R\,129. Therefore, rather than
making uncertain assumptions, we will derive the optical and NIR
extinction properties of R\,129 simply by dividing the average flux of 
the model spectrum inside each photometric band by the corresponding
values measured for this star. We will show in this way that also for
R\,129 the extinction properties are very similar to those
characteristic of 30\,Dor.

To verify that this approach gives meaningful results, we have applied
it also to R\,139. To achieve this, we derived an extinction curve from
the  comparison of the visible and NIR photometry with the model
spectrum in Figure\,\ref{fig1} and compared it with the extinction curve
measured in  the central regions of 30\,Dor (De Marchi \& Panagia 2014;
De Marchi et  al. 2016). As we will show in the next section, the
extinction curve derived photometrically for R\,139 is in excellent
agreement with the absolute extinction law of 30\,Dor.

\begin{figure*}
\centering
\resizebox{0.45\hsize}{!}{\includegraphics{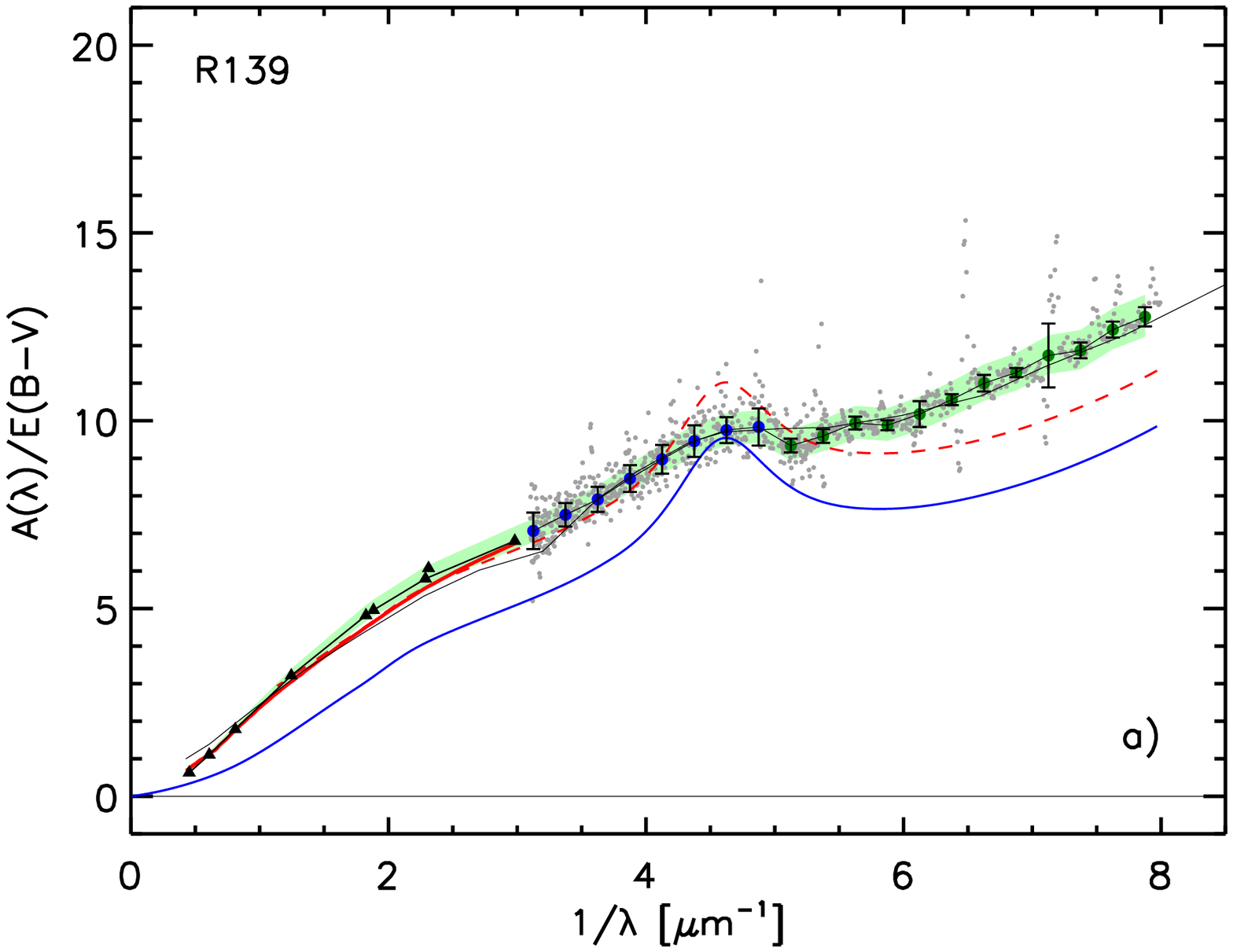}}
\resizebox{0.45\hsize}{!}{\includegraphics{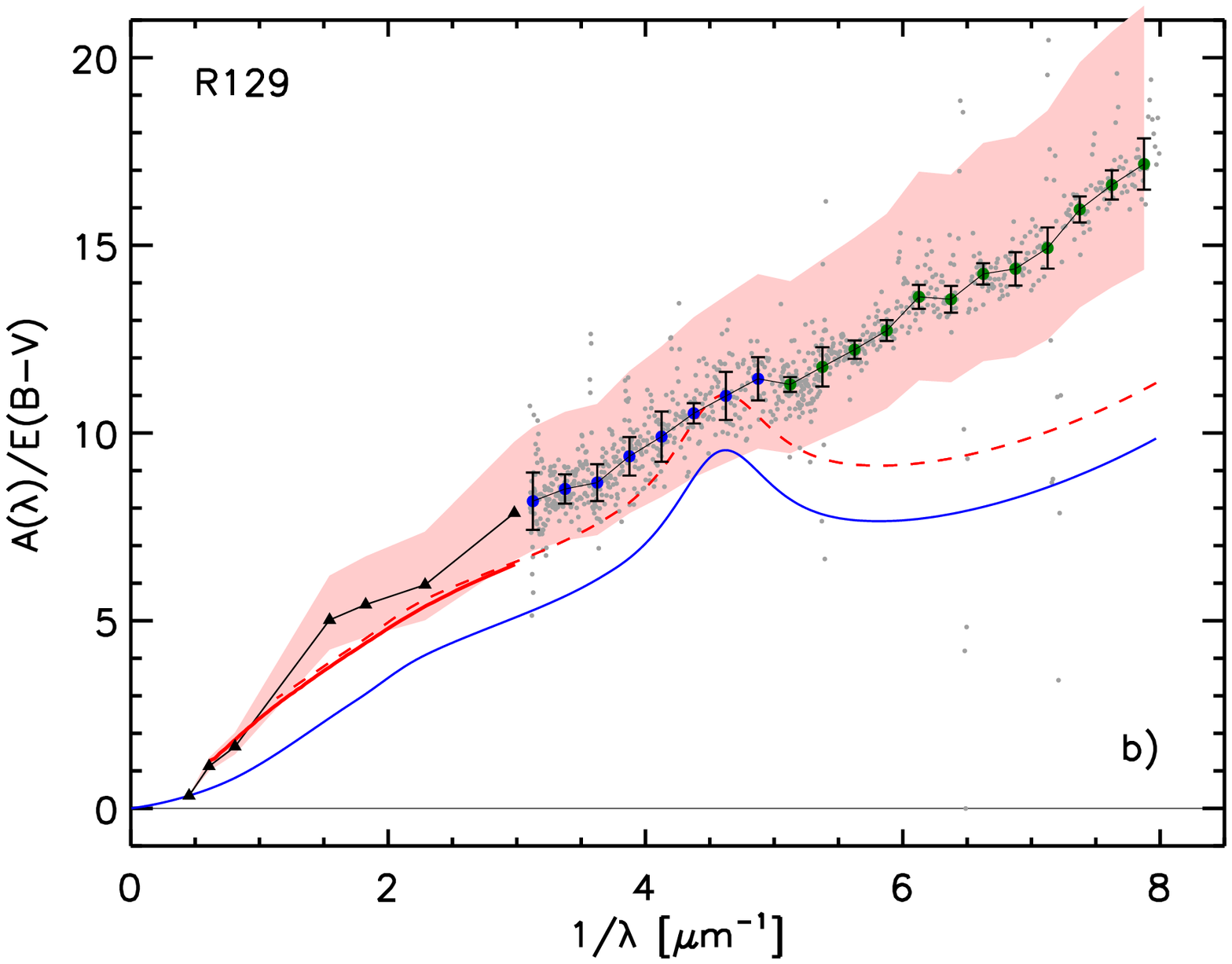}}
\caption{Measured extinction curves for R\,139 (Panel a) and R\,129
(Panel b), where the $A_\lambda$ values are shown as a function of
$1/\lambda$, normalised  by $E(B-V)$. The small grey dots are the result
of applying Equation\,\ref{eq1}, while the large dots show the average
values inside  $0.25$-wide bins in $1/\lambda$ and the error bars
indicate the associated  $\pm 1\,\sigma$ uncertainties on the mean. The
connected triangles indicate the empirical extinction curve
independently derived from the comparison of the visible and NIR
photometry with the model spectra in Figure\,\ref{fig1}. The shaded
areas show how the $\pm 1\,\sigma$ photometric uncertainties on  the
values of $B - V$ affect the normalisation of the extinction curves. For
reference, the thin blue solid lines correspond to the Galactic
extinction law, the dashed red curves are the same extinction law
modified with an additive factor of $1.5$ at optical wavelengths and a
multiplicative factor of 2 longwards of 1$\mu$m, and the solid red lines
indicate the optical and NIR extinction law of 30\,Dor. 
} 
\label{fig2}
\end{figure*}

\begin{figure*}
\centering
\resizebox{0.45\hsize}{!}{\includegraphics{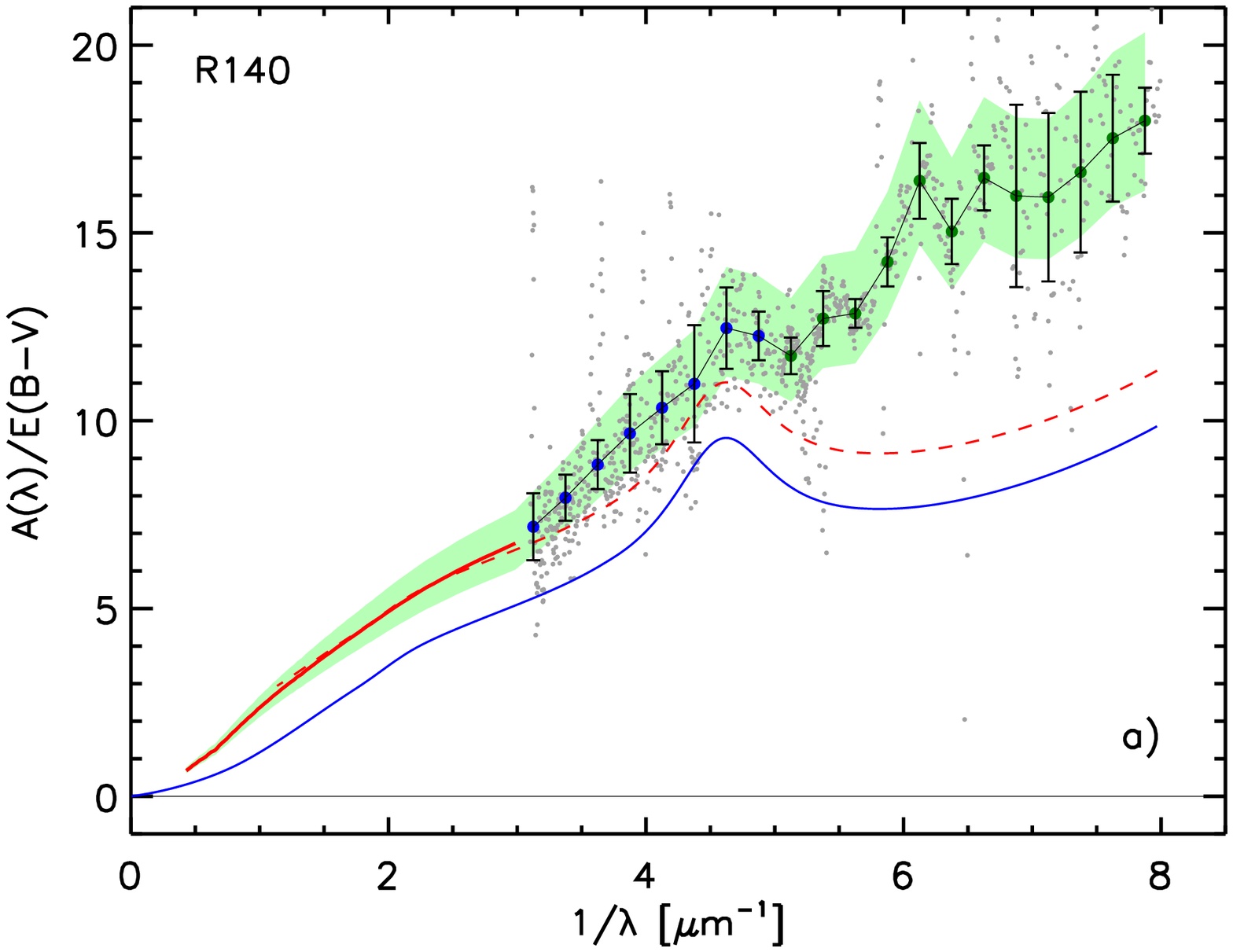}}
\resizebox{0.45\hsize}{!}{\includegraphics{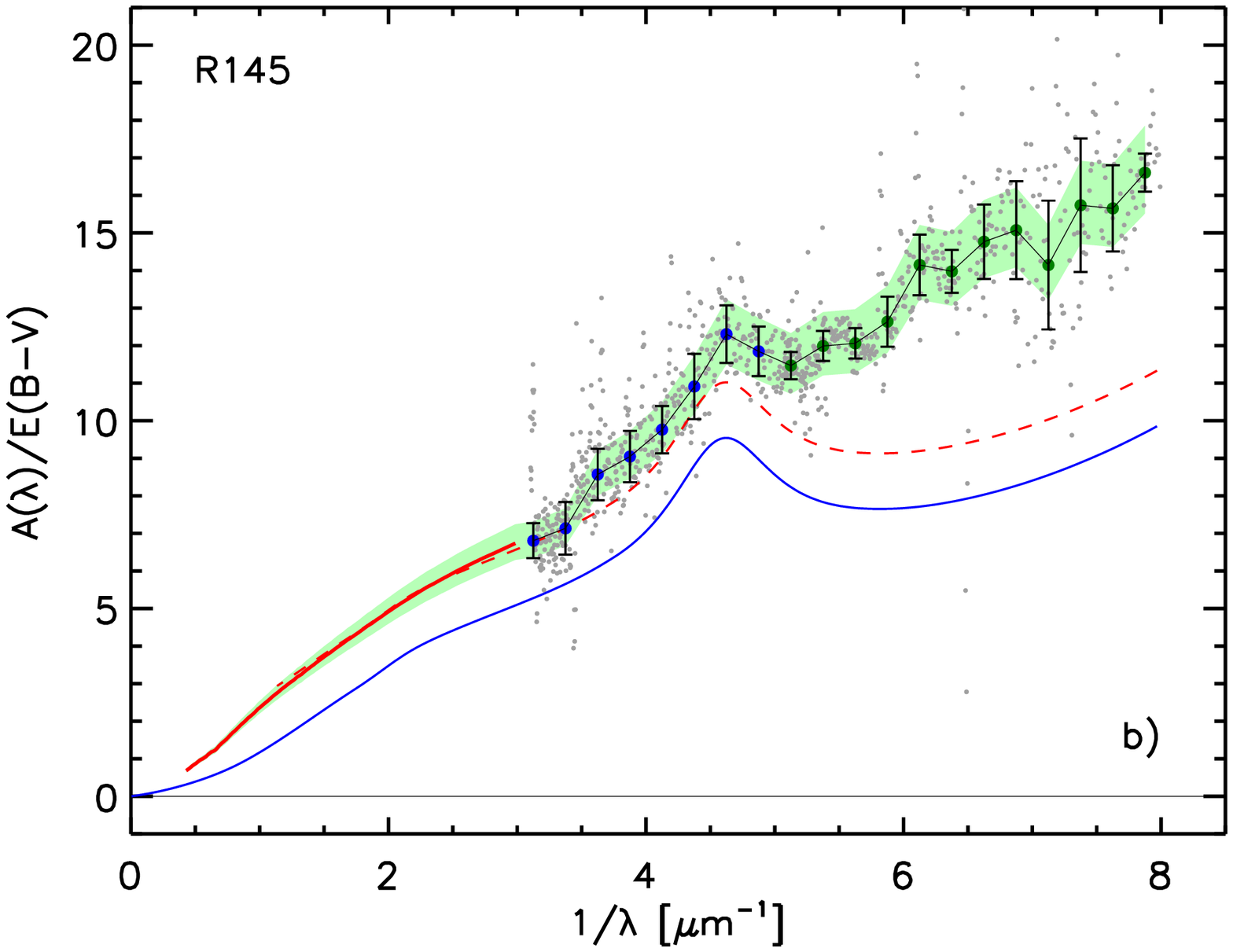}}
\caption{Same as Figure\,\ref{fig2} but for the WR stars R\,140 (Panel
a) and R\,145 (Panel b). } 
\label{fig3}
\end{figure*}

\section{Extinction curves}

In Figures\,\ref{fig2} and \ref{fig3} we show the extinction properties
derived for the four stars in our sample. The $A_\lambda$ values are
shown as a function of the wave number $1/\lambda$ and, as customary,
they are normalised by $E(B-V)$ (see Table\,\ref{tab1}), i.e.
$R(\lambda) \equiv A(\lambda)/E(B-V)$. We note that, unless otherwise
indicated, throughout this paper the values of $A(\lambda)$ and
$R(\lambda)$ referring to the four stars in our sample never include the
effects of the intervening MW absorption along the line of sight. 

At UV wavelengths, the small grey dots are the result of the
individual point-by-point ratios between spectra as per
Equation\,\ref{eq1}, while the large dots with error bars show the
average values inside $0.25$-wide bins in $1/\lambda$. The average is
calculated iteratively after applying a $2.5\,\sigma$ clipping to the
data in order to remove outliers. The final $\pm 1\,\sigma$
uncertainties on the mean are shown by the error bars. 

For reference, the thin blue solid lines correspond to the Galactic
extinction law (Fitzpatrick \& Massa 1990), while the dashed red curves
are the same extinction law modified to account for the effects of a
grey component as observed in 30\,Dor by De Marchi \& Panagia (2014).
This modification of the Galactic law consists in an additive factor of
$1.5$ at optical wavelengths and a multiplicative factor of 2 longwards
of 1$\mu$m (see De Marchi \& Panagia 2014 for details). The shaded areas
show how the $\pm 1\,\sigma$ photometric uncertainties on the values of
$B-V$ affect the normalisation of the extinction curves. In the UV, the 
resulting uncertainty band is generally compatible with the $\pm
1\,\sigma$ error bars described above, but in the case of R\,129 it is
almost four times as wide. This is due to the relatively large $\sim
0.05$\,mag uncertainty on $B-V$ combined with the  intrinsically small
value of $E(B-V)=0.20$. Other marginal sources of uncertainty on the
derived extinction curves are associated with fluctuations on the value
of the intervening MW extinction and possible spectral mismatches
between the observed stars and the adopted spectral models. In fact,
both are small. The former is the same for all objects in our field and
amounts to $0.01$\,mag (Brunet 1975; Fitzpatrick \& Savage 1984). The
latter is discussed in Section 3 and  amounts to a few percent at
most.    

At longer wavelengths we show as a thick solid red line the optical and
NIR  extinction law of 30\,Dor from De Marchi \& Panagia (2014) and De
Marchi et al. (2016), which we adopted in the study of the extinction
towards R\,139, R\,140, and R\,145. As mentioned earlier, for R\,139 we
also independently derived an empirical extinction curve from the
comparison of the visible and NIR photometry with the model spectrum in
Figure\,\ref{fig1}. The absolute normalisation of the curve is based on
the K-band value of $R_\lambda = A_\lambda/E(B-V)=0.78$, as measured by
De  Marchi \& Panagia (2014) in 30\,Dor. The  resulting curve, indicated
by the connected triangles in Figure\,\ref{fig2}a, is in excellent
agreement with the optical and NIR 30\,Dor extinction law (thick solid
red line).

Proceeding in the same way, we derived an empirical extinction curve
also for R\,129 in the NIR and optical range, also indicated by
connected triangles (see Figure\,\ref{fig2}b). The resulting curve is
similar to that of 30\,Dor and suggests a slightly larger value of
$R_V$, namely $5.3\pm0.07$ instead of $4.5\pm0.02$, revealing that also
in this region outside of 30\,Dor (180\,pc NNW of R\,136) there is
an excess of big grains compared to the diffuse Galactic ISM. Note that,
to be conservative, the empirical extinction curve of R\,129  has been
normalised to $R_K = 0.34$, namely the value characteristic for the
diffuse Galactic ISM (Fitzpatrick \& Massa 1990). If instead we had
adopted a $R_K$ value intermediate between those of the Galaxy and of
30\,Dor, namely $R_K=0.66 \pm 0.22$, the entire R\,129 curve in
Figure\,\ref{fig2}b would be shifted up solidly by the corresponding
amount, resulting in $R_V\simeq 5.5 \pm 0.2$. This is not surprising,
considering the value of $R_V=5.6 \pm 0.3$ measured by De Marchi et al.
(2014) in a region located about $6^\prime$ SW of 30\,Dor. 

The actual shape of the extinction law is linked to the composition of
the grains and to the distribution function of their sizes (e.g.
Greenberg 1968). In particular, selective depletion of small grains from
the mix has been shown to produce a flattening (weakening) of the
extinction curve at UV wavelengths (e.g. Mathis, Rumpl, Nordsieck 1977).
For example, the progressively flatter UV extinction curves observed
towards the nearby massive stars $\sigma$\,Sco, $\rho$\,Oph, and
$\theta^1$\,Ori are attributed by Mathis \& Wallenhorst (1981) to the
progressive disappearing of the small silicate grains first, followed by
that of the smallest graphite grains. 

Figures\,\ref{fig2} and \ref{fig3} immediately reveal that all four
extinction curves climb steeply into the FUV, more steeply than the 
Galactic extinction law at the same wavelengths. Qualitatively, we can
already conclude that there is a grain-size distribution function in
which small particles are at least as abundant as in the diffuse
Galactic ISM, both inside and outside 30\,Dor. 

For a more quantitative analysis, we compare the extinction curves of
R\,139 and R\,129. Both have late O spectral type and the continuum in
the spectra of stars of this type can be reliably modeled and reproduced
without having to worry in detail about the effects of mass loss and the
resulting stellar winds. The latter are important for the WR stars in
our sample and are more difficult to model (e.g. Bonanos et al. 2009).
Therefore, we focus our analysis on the O-type stars.

As Figure\,\ref{fig2} illustrates, shortwards of $\sim 1\,700$\,\AA, 
($1/\lambda > 6$) the extinction curves of R\,129 and R\,139 are almost
parallel to one another. The best fits to the slope of the curves in the
range $6<1/\lambda<8$ are, respectively, $1.43 \pm 0.11$ and $2.14 \pm
0.16$ (1\,$\sigma$). Both are steeper than the Galactic extinction law
in the same range ($1.12 \pm 0.03$). As mentioned above, this
immediately suggests a relative excess of small grains. 

It is well known ({\em e.g.} van de Hulst 1957; Greenberg 1968;
Draine \& Lee 1984) that at wavelengths short enough the extinction
cross section (absorption\,+\,scattering) of a grain tends
asymptotically to twice its geometric cross section $\sigma_{\rm geom} =
\pi \, a^2$, where $a$ denotes the grain radius. At longer wavelengths
the extinction is essentially pure absorption and the cross section is
smaller than $\sigma_{\rm geom}$, being proportional to $\sigma_{\rm
geom} \times 2\,\pi\,a / \lambda$. Conveniently enough, the
transition occurs approximately at $\lambda_0 \sim 2\,\pi\,a$. Thus, for
a fixed grain size, to a zeroth approximation one would expect a marked
change of slope, with the transition occurring around $\lambda_0$ and a
flattening at shorter wavelengths. When a more realistic distribution of
grain sizes is present, the transition will be obviously smoother, but
the extinction curve would flatten out only at wavelengths shorter than
$\lambda_0$. Since the extinction curves of both R\,139 and R\,129
continue to grow steeply with a steady slope and no abatement until the
low wavelength end of our data ($\lambda_{\rm min} \simeq 0.125\,\mu$m),
it appears that $\lambda_0$ is not yet reached. Thus it is safe to
assume that $\lambda_{\rm min} > \lambda_0$. In other words, the size of
the smallest grains must be smaller than that corresponding to
$\lambda_{\rm min}$, i.e. smaller than $\lambda_{\rm min} / (2\,\pi)$.
This suggests grain sizes of the order of $a \sim 0.01 - 0.02\,\mu$m. 

Even though a detailed study of the grain size distribution in 30\,Dor
is beyond the scope of this paper, because no information is available
on the nature of the absorbers, comparing the extinction curves of
our 30\,Dor sources with those of other massive stars in the LMC can
give us some indications about the small-grain population also in and
around 30\,Dor. A relevant comparison object is Sk\,--\,70\,116, a star
of spectral type  B2I located in LH\,117, an OB association (Massey et
al. 1989) about $1\fdg25$\,SE of 30\,Dor and included in the LMC\,2
Supershell sample studied by Gordon et al. (2003). 

\begin{figure}
\centering
\resizebox{0.9\hsize}{!}{\includegraphics{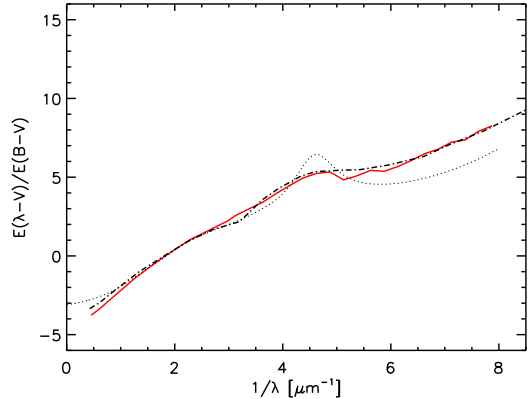}}
\caption{A different normalisation of the extinction curves to remove
the effects of grey extinction. The red solid curve corresponds to 
R\,139 and the dot-dashed line to Sk\,--\,70\,116. For comparison, the
thin dotted line is the MW extinction law.  } 
\label{fig4}
\end{figure}

To ease the comparison, in Figure\,\ref{fig4} we show the extinction
curves with a different normalisation, namely $E(\lambda -V)/E(B-V)$ in
turn equivalent to $(A_\lambda-A_V)/E(B-V)$ or $A_\lambda/E(B-V)-R_V$.
This normalisation is immune from the effects of the additive grey
extinction caused by an excess of large grains in the 30\,Dor region (De
Marchi \& Panagia 2014). The red solid curve corresponds to R\,139,
while the dot-dashed line to Sk\,--\,70\,116 (Gordon et al. 2003). The
agreement is excellent and applies to the entire wavelength range
spanned by the observations, including the FUV rise, the shape of the
characteristic $2\,175$\,\AA\ feature and the decline in the infrared.
We note in passing that the similarity in the shape of the latter
suggests that the $R_V=3.41\pm0.27$ quoted by Gordon et al. (2003) is
probably a lower limit to the actual value, which might be in excess of
$4$, also considering the uncertainties on the 2\,MASS photometry for
this object. Regardless of the actual value of $R_V$, the excellent
match between the curves reveals that the extinction properties towards
R\,139 are not unusual in this region of the LMC and confirms that even
inside 30\,Dor there is an excess of small grains with respect to the
Galactic extinction law. .

Also R\,140 and R\,145 confirm these conclusions (see
Figure\,\ref{fig3}). Although spectral comparison with the atmosphere of
WR stars is complicated by the presence of winds, that are not always
properly modelled (Bonanos et al. 2009), these two stars provide a
picture consistent with that outlined above. In fact, also two
supergiants of spectral type B included in the sample studied by Gordon
et al. (2003) in the LMC (Sk--67\,2 and Sk--69\,213) have extinction
curves in rather good agreement with those of R\,140 and R\,145,
including the shape of the 2\,175\,\AA\ feature. This shows, once more,
that the extinction conditions probed by R\,140 and R\,145 are similar 
to those found elsewhere in the LMC.

\begin{figure}
\centering
\resizebox{0.9\hsize}{!}{\includegraphics{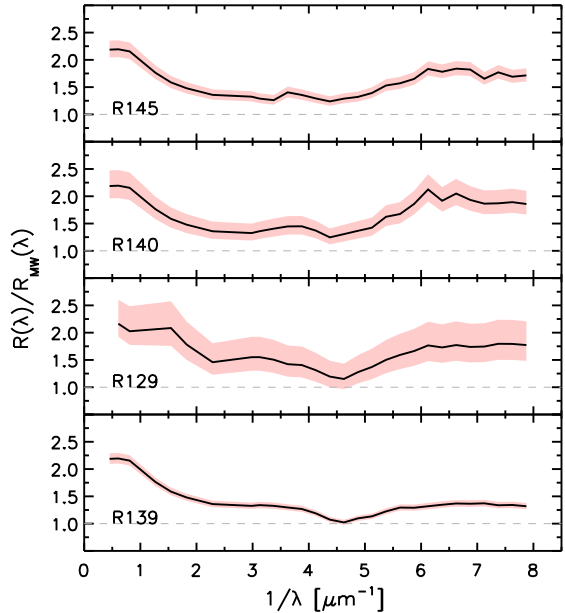}}
\caption{Ratio between the extinction curves of our 30 Dor stars and
that of the  Galactic ISM, normalised to $E(B-V)$. The shaded
areas show how the $\pm 1\,\sigma$ photometric uncertainties on  the
values of $B - V$ affect the normalisation of the extinction curves (see
Figures\,\ref{fig2} and \ref{fig3}).} 
\label{fig5}
\end{figure}

\section{Discussion}

In De Marchi \& Panagia (2014) we showed that, at optical wavelengths,
the  shape of the extinction curve of 30\,Dor has an additional grey 
component compared to the extinction law in the diffuse Galactic ISM. 
This reveals that in 30\,Dor there is a larger fraction of  big
grains relatively to the MW (see also De Marchi et al. 2016). These are
grains with  sizes of $0.05\,\mu$m $\la a \la 0.25\,\mu$m (hereafter,
for simplicity,  ``big grains''), whose effect is most directly seen on
the extinction curve in the  range $0.3 - 1.6\,\mu$m. To determine how
large this fraction of big grains actually is, we looked in the NIR,
where the wavelength is appreciably  greater than the size of all
grains, so that the extinction is dominated by  grains of larger size
and it is proportional to their total mass (e.g. Greenberg 1968). 

We observed that at wavelengths longer than 1\,$\mu$m, the extinction
curve of 30\,Dor tapers off as $\sim \lambda^{-1.7}$. This behaviour is
virtually identical to that of the Galactic extinction law (e.g.
Cardelli, Clayton \& Mathis 1989; Wang et al. 2013) and suggests that
the big grains in 30\,Dor have the same nature as those in the Galaxy.
The ratio between the 30\,Dor and Galactic curves is $\sim2$. {Thus,
even  though we do not have direct information on the nature of the big
grains  in 30\,Dor, the most reasonable explanation for the extinction
curve that  we observe at wavelengths longer than 1\,$\mu$m is that the
nature of big  grains in 30\,Dor is the same as in the diffuse ISM of
the Galaxy or LMC,  but that the fraction of these big grains is about
twice as high compared  with the grain population in the Milky Way.}

The analysis presented in the previous section, and in particular
the  steep FUV rise in Figures\,\ref{fig2} and \ref{fig3}, indicates
that  in 30\,Dor also grains with $0.01\,\mu$m $\la a \la 0.02\,\mu$m 
(hereafter, for simplicity, ``small grains'') are overabundant compared
to  the diffuse Galactic ISM.\footnote{In the following discussion we
will use as a reference the diffuse Galactic ISM, but we point out that
the latter is in excellent agreement with the diffuse ISM in the LMC
(e.g. Gordon et al. 2003).} To gain some quantitative information on the
excess of small grains, we consider in Figure\,\ref{fig5} the ratio
between the extinction curves of our 30\,Dor stars $R_{30D}(\lambda)$ 
and that of the Galactic ISM, $R_{MW}(\lambda)$, taken from Fitzpatrick
\& Massa (1990). We recall that $R_{30D}(\lambda)$ does not include the
intervening MW absorption along the line of sight to the LMC.

As mentioned above, at NIR wavelengths ($1/\lambda \lesssim 1$) the
ratio of extinction curves is constant for all stars and has a value of
about 2, indicating that the nature of the big grains is the same and
their fraction is about twice as high as in the MW. A rather constant
ratio is also seen in the FUV ($1/\lambda \gtrsim 6$), where
$R_{30D}/R_{MW}\simeq 1.5$ for the O-type stars, and this suggests that
also the nature of the small grains is the same as in the MW and that in
30\,Dor their relative abundance is about 50\,\% higher than in the
diffuse  Galactic ISM. We stress that, even though small grains of a
different  nature could still reproduce the observed extinction curve
with a  different relative abundance, the most reasonable explanation is
that the nature of the small grains is the same as in the MW. The
observations of  the WR stars R\,140 and R\,145 suggest a higher ratio,
closer to $\sim 2$,  but this might be partly due to the strong winds of
these WR stars, as mentioned before, rather than to the intrinsic
properties of the ISM. Thus, we can safely conclude that the increase in
the abundance of small grains is at least 50\,\% compared to the MW, and
possibly larger. 

The other interesting feature of Figure\,\ref{fig5} is the marked
minimum near $1/\lambda \simeq 4.5$, corresponding to the
characteristic  $2\,175$\,\AA\ feature in the extinction curves. For the
O-type stars (R\,129 and R\,139), the ratio at this wavelength is
compatible with unity, and even for the WR stars (R\,140, R\,145) it is
only slightly higher ($\sim 1.2$). It appears that the processes 
responsible for the observed excess of both big grains and small silicate 
grains do not also contribute significantly to the very small 
carbonaceous particles with $a < 0.005\,\mu$m (hereafter, ``very small 
grains''), including possible graphitic grains and polycyclic aromatic 
hydrocarbons, which are traditionally associated with the $2\,175$\,\AA\ 
feature (e.g. Weingartner \& Draine 2001; Li \& Draine 2001; Bradley et 
al. 2005). According to Weingartner \& Draine (2001), the depletion of 
the $2\,175$\,\AA\ feature observed in dense environments, where $R_V$ is 
large like in this case, might indeed result entirely from the lack of 
very small carbonaceous grains.

As we already pointed out in the Introduction, a larger fraction of big 
grains requires a mechanism that either selectively adds big grains, or
selectively removes small grains by destroying or co-adding them, or a
combination of both.  However, any mechanism that produces big 
grains only by selectively removing or co-adding pre-existing small 
grains is not compatible with the observations, since not only the big 
grains, but also the small grains appear to be more abundant in 30 Dor 
than in the diffuse ISM of either the Galaxy or LMC. Therefore, Occam's 
razor would suggest that the injection of fresh grains (both big and
small) by 30\,Dor into the LMC diffuse ISM is the preferred mechanism.

The resulting picture is one in which we identify two components in the 
ISM of 30\,Dor. One is similar to that of the diffuse Galactic ISM and
most likely represents the pre-existing diffuse ISM of the LMC (Gordon
et al. 2003). The additional, newly injected component contains 
big grains with a relative abundance similar to that of the ISM in the
MW or LMC, small grains with a relative abundance of about half
that of the  MW/LMC, and little or no very small carbonaceous grains.
Thus, the distribution of the total mass stored in grains of different
sizes in the newly injected component appears to be steeper than that of
the diffuse ISM in the Galaxy or LMC, i.e. skewed towards larger values
of the radius $a$.

A process able to naturally account for the injection of new grains in a
young region where star formation is still ongoing is the explosion of
massive stars as type-II supernovae (SN). Indeed, massive-star SN are
expected to have been major dust producers throughout the history of 
the universe (e.g. Sugerman et al. 2006). Observations of SN\,1987A
with Herschel and ALMA (Matsuura et al. 2011;  Gomez et al. 2012;
Indebetouw et al. 2014) and of the Crab Nebula and  Cassiopeia A with
Herschel  (Barlow et al. 2010; Gomez et al. 2012; De Looze et al. 2017)
have revealed up to and possibly exceeding $\sim 0.5$\,\Msolar of dust
formed in situ in core-collapse SN ejecta. 

To be sure, there is still considerable uncertainty about the actual
size distribution of dust formed in supernova remnants (e.g. Temim \&
Dwek 2013; Wesson et  al. 2015; Owen \& Barlow 2015; Bevan \& Barlow
2016). Nevertheless, Gall et al. (2014) discovered rapid formation
(within hundreds of days) of dust  in the dense circumstellar medium of
SN\,2010jl and showed that, adopting a  power-law distribution of grain
sizes with a typical index $\sim 3.2 <  \alpha < 3.8$, the observed
extinction curves can be reproduced only if there is an excess of large
grains and the maximum size of the grains exceeds  $a>1\,\mu$m.
Therefore, within the stated uncertainties, big  grains appear to
represent the majority of the produced dust mass. Simulations by Silvia
et al. (2010) indicate that  grains larger than $\sim 0.1\,\mu$m survive
reverse shock interactions  and only a low fraction of them is sputtered
to smaller radii. Similarly,  Biscaro \& Cherchneff (2016) show that in
SN with dense ejecta, like  SN\,1987A, also grains down to $0.05\,\mu$m
survive thermal sputtering in  the remnant. Therefore, the big grains to
which our observations are sensitive ($a > 0.05\,\mu$m) appear to be a
natural and long-lasting  end-product of type-II SN explosions. 

Note that there is currently no way to know whether these  big
grains are produced directly as such by SN explosions or whether they 
result from quick growth of smaller grains produced in the ejecta. But
for the purposes of our research, which is to understand the extinction 
properties in massive star-forming regions, this is not important. What
is  clear, however, is that even if the big grains were forming through
the quick  growth of small grains, the latter cannot be just the
pre-existing small grains in the ISM of the LMC, because in the 30\,Dor
ISM their relative amount is about 50\,\% higher than in the diffuse ISM
of the LMC or MW. In fact, unlike big grains, small grains may be
destroyed in the shocked  region within the type-II SN before being
injected into the interstellar space (Bianchi \& Schneider 2007; Nozawa
et al. 2007) and might have been originally even more abundant.
Therefore, pre-existing small grains  alone and any process that they
might have undergone cannot account for the spectrum of grain sizes and
masses that our observed extinction curves reveal. Instead, fresh
injection of new grains is necessary, and the spectrum of grain sizes
for dust effectively injected by type-II SN explosions into the
surrounding ISM agrees, at least qualitatively, with the mass
distribution that we infer from the observations. 

To be more quantitative, we compare the total mass of big grains implied
by the extinction curve that we measure with that expected from SN-II
explosions in the Tarantula Nebula. At wavelengths longer than the $V$
band, the effects of big grains start to saturate, suggesting that
$\lambda_0 \simeq 0.55\,\mu$m (see Section\,4) and, consequently, the
presence of grains of radius $a \simeq 0.09\,\mu$m or $a \simeq 9 \times
10^{-6}$\,cm. As noted in Section\,4, at wavelengths short enough, the 
extinction cross-section $\sigma_{\rm ext}$ of a grain of radius $a$
tends asymptotically to $ 2 \, \pi \, a^2$, corresponding to $5.1 \times
10^{-10}$\,cm$^{-2}$ in this case. Knowing the grain's cross section, we
can derive the number of grains along the line of sight simply as a
ratio of optical depth $\tau_v$ and $\sigma_{\rm ext}$.  

The median value of $A_V$ in the field of the Tarantula Nebula is
$A_V=1.45$ (De Marchi et al. 2016), about 1/3 of which is due to the
freshly injected component of big grains, since $R_V=4.5$ in the
Tarantula Nebula and $R_V=3.1$ in the diffuse ISM of the LMC. The
resulting $A_V$ value associated with the freshly injected big grains
alone is then $A_V= 0.45$, which in turn corresponds to an optical depth
$\tau_V = 1.086 \times A_V = 0.49$. The number of big grains associated
with the additional component along the line of sight is simply 

\begin{equation}
n = \tau_V / \sigma_{\rm ext} = 9.6 \times 10^8. 
\end{equation}

\noindent
Assuming a spherical shape for the Tarantula Nebula, with
diameter $L=200\,pc=6.17 \times 10^{20}$\,cm (Sabbi et al. 2013), the 
spatial density of big grains along the line of sight is 

\begin{equation}
\rho = n / L = 1.56 \times 10^{-12}\,{\rm g\,cm}^{-3}.
\end{equation}

\noindent
With a total volume $V=1.23 \times 10^{62}$\,cm$^3$, the total number of
new big grains hosted in the Tarantula Nebula is 

\begin{equation}
N=\rho \times V=1.9 \times 10^{50}.
\end{equation}

\noindent
Silicates grains have a specific density of $\sim 2.7$\,g\,cm$^{-3}$
(Panagia 1974; Draine \& Lee 1984), implying a typical mass per grain of
$m=8.2 \times 10^{-15}$\,g, and a total mass of freshly injected big
grains of 

\begin{equation}
M= N \times m = 1.9 \times 10^{50} \times  8.2 \times 10^{-15} = 1.6
\times 10^{36}\,{\rm g},
\end{equation}

\noindent
or about $800$\,\Msolar.

If we assume that each SN-II explosion provides about $0.3$\,\Msolar of
big grains (see above), then we need about 2400 such explosion occurring
in the Tarantula Nebula over the time of its existence. Recently,
Cignoni et al. (2015, 2016) studied the properties of the stellar
populations present in the Tarantula region, using synthetic
colour--magnitude diagrams generated with the latest PARSEC models
(Bressan et al. 2012; Marigo et al. 2017) and assuming a Kroupa (2001)
initial mass function (IMF; $\gamma=2.3$). They showed that star
formation within the whole Tarantula Nebula has been active for more
that  20\,Myr. Also, for one of the more evolved clusters in the
Tarantula region, Hodge\,301, they estimate an age of about 29\,Myr, a
total stellar mass of 8800\,Msolar, and a number of about 49 SN-II
explosions occurred since its birth. We adopt as a conservative value
for the total mass of the Tarantula $10^6$\,\Msolar, considering that
the R\,136 cluster alone has a mass of $4.5 \times 10^5$\,\Msolar with a
half-mass radius of 14\,pc  (Bosch et al. 2009). Thus, using the data of
Hodge\,301 to base our extrapolation, we obtain about 5600 as a lower
limit to the total number of SN-II occurred since the Tarantula Nebula
formation (i.e. no less than of 30\,Myr ago). In fact, the rate of SN-II
could be $\sim 70\,\%$ higher adopting the ``shallower'' IMF index
$\gamma=1.9$, as recently determined by Schneider et al. (2018) for
30\,Dor. Therefore, although these calculations are necessarily
approximate, it appears that the total mass of big grains required by
our extinction curves is quantitatively consistent with that expected
from SN-II explosions.

\section{Conclusions}

Our work suggests that type-II SN explosions can modify the
conditions of the ISM by injecting a mixture of grains with a different
distribution of sizes. This change is only temporary, because grain
shattering will eventually affect the newly injected grains, breaking
bigger grains, and shocks will ultimately destroy them and increase the
abundance of small grains (e.g. Jones et al. 1996; Hirashita \& Kuo
2011). 

The lifetime of dust grains in the Magellanic Clouds, as calculated by
Temim et al. (2015) from the analysis of the complete sample of SN
remnants in these galaxies, ranges from $\sim 20$ to $\sim 70$\,Myr,
depending on their composition and location. Considering that the
lifetime of the lowest-mass type-II SN progenitors ($\sim 8$\,\Msolar)
is of the order of 40\,Myr and for the most massive stars only a few Myr
(e.g. Marigo et al. 2017), the modified distribution of grain sizes is
likely to affect the ISM for about $\sim 50 - 100$\,Myr starting some
$\sim 10$\,Myr after a major star-formation event, also including the
time needed for the ejecta from the first SN explosions to be dispersed
in the surrounding ISM. Obviously, regions of extended and
protracted star formation like 30\,Dor and the entire Tarantula Nebula
(De Marchi et al. 2017; Sabbi et al. 2016; Cignoni et al. 2015) will
bear the consequences  of a modified ISM for a longer time.  Indeed, in
the somewhat older neighbouring star-forming region located about
$6^\prime$ SW of 30\,Dor, De Marchi et al. (2014) measured an even
higher value of $R_V$, namely $5.6 \pm 0.3$, indicating an even larger
fraction of big grains. Nor is the contamination of the ISM with freshly
injected grains limited only to the immediate surroundings of the
star-forming regions themselves. 

Assuming a typical velocity dispersion of $\sim 5 - 10$\,km\,s$^{-1}$ in
massive star-forming regions (e.g. Bosch, Terlevich \& Terlevich 2009;
Rochau et al. 2010) as a very conservative lower limit to the velocity
of the ejecta in the ISM, some $\sim 50$\,Myr after a major
star-formation event we should expect the freshly created grains to have
travelled distances of $\sim 500$\,pc, as long as they are not slowed
down or destroyed in their flight.  

 Thus, in star-forming regions hosting stars that end their life as
SN-II, and hence more massive than $\sim 8$\,\Msolar,  we should
expect extinction properties appreciably different from those of the
diffuse ISM. This includes the nearby Orion Nebula Cluster, whose
anomalous extinction properties have been noticed first by Baade \&
Minkowski (1937) and for which a value of  $R_V$ in excess of $\sim 5$
has been consistently reported since  the 1950s (e.g. Sharpless 1952;
Johnson \& Borgman 1963; Johnson \& Mendoza 1964; Gebel 1968; Breger et
al. 1981). The observed extinction properties are attributed to an
additional component of big grains (e.g. Cardelli \& Clayton 1988;
Beitia--Antero \& Gomez de Castro 2017). These properties are fully
compatible with those observed in 30\,Dor and can be attributed to fresh
injection of preferentially big grains by type-II SN explosions. 

Conversely, star-forming regions of low total mass should not
display anomalous extinction curves because they do not contain enough
massive stars whose explosion as type-II SN can effectively alter the
conditions of the ISM. A nearby example is that of the Taurus
star-forming region, with a very low total mass ($\sim 50$\,\Msolar; De
Marchi et al. 2010) and membership (438 members; Luhman 2018), and with
extinction properties (Vrba \& Rydgren 1985; Kenyon et al. 1994; Arce \&
Goodman 1999) fully consistent with those of the diffuse Galactic ISM.

Our work highlights the complexity inherent in interpreting observations
of star-forming regions, even in the Milky Way and nearest Magellanic
Clouds galaxies. Deriving stellar masses and star formation rates from
broad-band photometry and emission-line luminosities in more distant
starburst galaxies and at high redshift is subject to even more severe
limitations, because the extinction properties and three-dimensional
distribution of the stars with respect to the dust are completely
unknown. Yet the only star-forming regions that can be observed in
these early Universe environments are clusters as massive as 30\,Dor or
more. They can only be observed when massive stars are still present,
and therefore when the ISM is mostly affected by big grains injected by
SN-II. Therefore, without proper knowledge of the actual extinction 
properties, any attempts to derive quantitative physical parameters in
these environments will be plagued by very large uncertainties and are
bound to fail.

\vspace*{0.05cm}
\begin{acknowledgements}

We are indebted to an anonymous referee, whose constructive criticism
has helped us to improve the presentation of this work. NP acknowledges
partial support by HST-NASA grants GO-11547.06A and GO-11653.12A, and
STScI-DDRF grant D0001.82435.

\end{acknowledgements}

\end{document}